\documentclass[%
aps,
superscriptaddress,
preprint,
showpacs,
prc,
]{revtex4-1}

\usepackage{graphicx}% Include figure files
\usepackage{dcolumn}% Align table columns on decimal point
\usepackage{bm}% bold math
\usepackage[mathlines]{lineno}% Enable numbering of text and display math

\usepackage{lineno}

\begin{document}

\preprint{APS/123-QED}

\title{Differential cross section and photon-beam asymmetry for 
the $\vec{\gamma} p$ $\rightarrow$ $\pi^{+}n$ reaction at forward 
$\pi^{+}$ angles for $E_{\gamma}$=1.5-2.95 GeV}% Force line breaks with \\

\newcommand{\KUADDRESS}{Department of Physics, Korea University, Seoul 02841, Republic of Korea}
\newcommand{\APCTPADDRESS}{Asia Pacific Center for Theoretical Physics, Pohang, Gyeongbuk, 37673, Republic of Korea}
\newcommand{\IBSADDRESS}{Rare Isotope Science Project, Institute for Basic Science, Daejeon 34047, Korea}
\newcommand{\OHIOADDRESS}{Department of Physics and Astronomy, Ohio University, Athens, Ohio 45701, USA}
\newcommand{\KYOTOADDRESS}{Department of Physics, Kyoto University, Kyoto 606-8502, Japan}
\newcommand{\KYOTOOCRADDRESS}{Office of Commnunity Relations, Faculty of Science, Kyoto University, Kyoto 606-8502, Japan}
\newcommand{\KONANADDRESS}{Department of Physics, Konan University, Kobe, Hyogo 658-8501, Japan}
\newcommand{\XFELADDRESS}{XFEL Project Head Office, RIKEN, Sayo, Hyogo 679-5143, Japan}
\newcommand{\CENTADDRESS}{Department of Physics, National Central University, Taoyuan City 32001, Taiwan}
\newcommand{\SINICAADDRESS}{Institute of Physics, Academia Sinica, Taipei 11529, Taiwan }
\newcommand{\CHEMADDRESS}{ChemMatCARS, The University of Chicago, Argonne, Illinois 60439, USA}
\newcommand{\NSRRCADDRESS}{Light Source Division, National Synchrotron Radiation Research Center, Hsinchu, 30076, Taiwan }
\newcommand{\RCNPADDRESS}{Research Center for Nuclear Physics, Osaka University, Ibaraki, Osaka 567-0047, Japan}
\newcommand{\JASRIADDRESS}{Japan Synchrotron Radiation Research Institute, Sayo, Hyogo 679-5143, Japan}
\newcommand{\YAMAGATAADDRESS}{Department of Physics, Yamagata University, Yamagata 990-8560, Japan}
\newcommand{\NAGOYAADDRESS}{Kobayashi-Maskawa Institute, Nagoya University, Nagoya, Aichi 464-8602, Japan}
\newcommand{\TOHOKUADDRESS}{Research Center for Electron Photon Science, Tohoku University, Sendai, Miyagi 982-0826, Japan}
\newcommand{\CHIBAADDRESS}{Department of Physics, Chiba University, Chiba 263-8522, Japan}
\newcommand{\MIYAZAKIADDRESS}{Department of Applied Physics, Miyazaki University, Miyazaki 889-2192, Japan}
\newcommand{\OSAKAADDRESS}{Department of Physics, Osaka University, Toyonaka, Osaka 560-0043, Japan}
\newcommand{\TOKYOIADDRESS}{Department of Physics, Tokyo Institute of Technology, Tokyo 152-8551, Japan}
\newcommand{\GIFUADDRESS}{Department of Education, Gifu University, Gifu 501-1193, Japan}
\newcommand{\RIKENADDRESS}{RIKEN Nishina Center, 2-1 Hirosawa, Wako, Saitama 351-0198, Japan}
\newcommand{\PKNUADDRESS}{Department of Physics, Pukyong National University, Busan 48513, Republic of Korea}
\newcommand{\JINRADDRESS}{Joint Institute for Nuclear Research, Dubna, Moscow Region, 142281, Russia}
\newcommand{\MSUADDRESS}{National Superconducting Cyclotron Laboratory, Michigan State University, East Lansing, MI 48824, USA}
\newcommand{\WAKAYAMAADDRESS}{Wakayama Medical College, Wakayama, 641-8509, Japan}
\newcommand{\SASKAADDRESS}{Department of Physics and Engineering Physics, University of Saskatchewan, Saskatoon, SK S7N 5E2, Canada}
\newcommand{\KEKADDRESS}{High Energy Accelerator Organization (KEK), Tsukuba, Ibaraki 305-0801, Japan}
\newcommand{\MINESOTAADDRESS}{School of Physics and Astronomy, 
University of Minnesota, Minneapolis, MN 55455, USA}
\newcommand{\PROTEINADDRESS}{Institute for Protein Research, 
Osaka University, Suita, Osaka 565-0871, Japan}
\newcommand{\AKITAADDRESS}{Akita Research Institute of Brain and 
Blood Vessels, Akita 010-0874, Japan}
\newcommand{\NDAADDRESS}{Department of Applied Physics, 
National Defense Academy, Yokosuka, Kanagawa 239-8686, Japan}
\newcommand{\KAOADDRESS}{Department of Physics, National Kaohsiung Normal University, Kaohsiung 824, Taiwan}
\newcommand{\CONNEADDRESS}{Department of Physics, University of Connecticut, Storrs, CT 06269-3046, USA}
\newcommand{\FUKUIADDRESS}{Proton Therapy Center, Fukui Prefectural Hospital, Fukui 910-8526, Japan}
\newcommand{\JAEAADDRESS}{Advanced Science Research Center, 
Japan Atomic Energy Agency, Tokai, Ibaraki 319-1195, Japan}
\newcommand{\CHUNGCHENGADDRESS}{Department of Physics, 
National Chung Cheng University, Chiayi 62102, Taiwan}
\newcommand{\KRISSADDRESS}{Korea Research Institute of Standards 
and Science (KRISS), Daejeon 34113, Republic of Korea}
\newcommand{\GENKENADDRESS}{National Institutes for Quantum and Radiological Science and Technology, Tokai, Ibaraki 319-1195, Japan}
\newcommand{\TOKYOADDRESS}{Department of Radiology, The University of Tokyo Hospital, Tokyo 113-8655, Japan}
\newcommand{\CASADDRESS}{Institute of High Energy Physics, Chinese Academy of Sciences, Beijing 100049, China}
\newcommand{\MICHIGANADDRESS}{Physics Department, University of Michigan, Michigan 48109-1040, USA}
\newcommand{\CROSSADDRESS}{Neutron Science and Technology Center, Comprehensive Research Organization for Science and Society (CROSS), Tokai, Ibaraki 319-1106, Japan}

\author{H.~Kohri}\affiliation{\RCNPADDRESS}\affiliation{\SINICAADDRESS}
\author{S.~Y.~Wang}\affiliation{\SINICAADDRESS}\affiliation{\CHEMADDRESS}
\author{S.~H.~Shiu}\affiliation{\SINICAADDRESS}\affiliation{\CENTADDRESS}
\author{W.~C.~Chang}\affiliation{\SINICAADDRESS}
\author{Y.~Yanai}\affiliation{\RCNPADDRESS}
\author{D.~S.~Ahn}\affiliation{\RIKENADDRESS}
\author{J.~K.~Ahn}\affiliation{\KUADDRESS}
\author{J.~Y.~Chen}\affiliation{\NSRRCADDRESS}
\author{S.~Dat$\acute{\rm{e}}$}\affiliation{\JASRIADDRESS}
\author{H.~Ejiri}\affiliation{\RCNPADDRESS}
\author{H.~Fujimura}\affiliation{\WAKAYAMAADDRESS}
\author{M.~Fujiwara}\affiliation{\RCNPADDRESS}\affiliation{\GENKENADDRESS}
\author{S.~Fukui}\affiliation{\RCNPADDRESS}
\author{W.~Gohn}\affiliation{\CONNEADDRESS}
\author{K.~Hicks}\affiliation{\OHIOADDRESS}
\author{A.~Hosaka}\affiliation{\RCNPADDRESS}
\author{T.~Hotta}\affiliation{\RCNPADDRESS}
\author{S.~H.~Hwang}\affiliation{\KRISSADDRESS}
\author{K.~Imai}\affiliation{\JAEAADDRESS}
\author{T.~Ishikawa}\affiliation{\TOHOKUADDRESS}
\author{K.~Joo}\affiliation{\CONNEADDRESS}
\author{Y.~Kato}\affiliation{\NAGOYAADDRESS}
\author{S.~H.~Kim}\affiliation{\APCTPADDRESS}\affiliation{\RCNPADDRESS}
\author{Y.~Kon}\affiliation{\RCNPADDRESS}
\author{H.~S.~Lee}\affiliation{\IBSADDRESS}
\author{Y.~Maeda}\affiliation{\FUKUIADDRESS}
\author{T.~Mibe}\affiliation{\KEKADDRESS}
\author{M.~Miyabe}\affiliation{\TOHOKUADDRESS}
\author{Y.~Morino}\affiliation{\KEKADDRESS}
\author{N.~Muramatsu}\affiliation{\TOHOKUADDRESS}
\author{T.~Nakano}\affiliation{\RCNPADDRESS}
\author{Y.~Nakatsugawa}\affiliation{\KEKADDRESS}\affiliation{\CASADDRESS}
\author{M.~Niiyama}\affiliation{\KYOTOADDRESS}
\author{H.~Noumi}\affiliation{\RCNPADDRESS}
\author{Y.~Ohashi}\affiliation{\JASRIADDRESS}
\author{T.~Ohta}\affiliation{\RCNPADDRESS}\affiliation{\TOKYOADDRESS}
\author{M.~Oka}\affiliation{\RCNPADDRESS}
\author{J.~D.~Parker}\affiliation{\KYOTOADDRESS}\affiliation{\CROSSADDRESS}
\author{C.~Rangacharyulu}\affiliation{\SASKAADDRESS}
\author{S.~Y.~Ryu}\affiliation{\RCNPADDRESS}
\author{T.~Sawada}\affiliation{\SINICAADDRESS}\affiliation{\MICHIGANADDRESS}
\author{H.~Shimizu}\affiliation{\TOHOKUADDRESS}
\author{Y.~Sugaya}\affiliation{\RCNPADDRESS}
\author{M.~Sumihama}\affiliation{\GIFUADDRESS}
\author{T.~Tsunemi}\affiliation{\KYOTOADDRESS}
\author{M.~Uchida}\affiliation{\TOKYOIADDRESS}
\author{M.~Ungaro}\affiliation{\CONNEADDRESS}
\author{M.~Yosoi}\affiliation{\RCNPADDRESS}

\collaboration{LEPS Collaboration}%\noaffiliation

\date{\today}% It is always \today, today,

\begin{abstract}
Differential cross sections and photon-beam asymmetries for the 
$\vec{\gamma} p$ $\rightarrow$ $\pi^{+}n$ 
reaction have been measured for 0.6$<$ $\cos\theta_{\pi}$ $<$1 
and $E_{\gamma}$=1.5-2.95 GeV at SPring-8/LEPS. 
The cross sections monotonically decrease as the photon beam 
energy increases for 
0.6$<$ $\cos\theta_{\pi}$ $<$0.9. 
However, the energy dependence of the cross sections 
for 0.9$<$ $\cos\theta_{\pi}$ $<$1 and $E_{\gamma}$=1.5-2.2 GeV 
($W$=1.9-2.2 GeV) is different, which 
may be due to a nucleon or $\Delta$ resonance. 
The present cross sections agree well with the previous 
cross sections measured by other groups and 
show forward peaking, suggesting significant 
$t$-channel contributions in this kinematical region. 
The asymmetries are found to be positive, which can be explained 
by $\rho$-exchange in the $t$-channel. 
Large positive asymmetries in the small $|t|$ region, 
where the $\rho$-exchange contribution becomes small, 
could be explained by introducing $\pi$-exchange interference 
with the $s$-channel. 
\end{abstract}

\pacs{13.60.Le,14.20Gk,14.40Aq,14.70.Bh,25.20Lj}
%\pacs{13.60.Le, 13.60.Rj, 13.88.+e, 14.20.Jn, 25.20.Lj}

\maketitle

%\tableofcontents

\section{Introduction}

%%%%%% pi+ n
Photoproduction of mesons is of special importance in the search of 
missing nucleon resonances. 
In quark models, there exist more nucleon resonances than have been 
experimentally observed so far~\cite{Capstick}. 
Since the nucleon resonances have relatively wide widths and are 
overlapping in mass, it is necessary to establish new resonances 
by performing partial wave analyses based on rich physics observables
over wide angular and energy regions. 
The $\gamma p$ $\rightarrow$ $\pi^{+}n$ reaction is one 
of the most basic photoproduction reactions. 
The $\gamma p$ $\rightarrow$ $\pi^{+}n$ reaction has relatively
large cross sections of $\sim$10 $\mu$b, 
which enables measurements of physics observables to 
clarify the reaction dynamics. 
JLab/CLAS has measured differential cross sections~\cite{Dugger1} 
in a wide angular region without forward and backward $\pi^{+}$ angles 
for $E_{\gamma}$=0.725-2.875 GeV. 
Existing data at forward $\pi^{+}$ angles taken for 
$E_{\gamma}$=1.2-3.0 GeV by 
DESY~\cite{Boschhorn1,Boschhorn2} are scarce and inadequate
to search for missing resonances. 
The SPring-8/LEPS experiments, with a spectrometer at 
forward angles, are complementary to CLAS experiments and
can provide valuable data for the missing resonance search. 

%%%%% Photon-Beam Asymmetry
We measured differential cross sections and 
photon-beam asymmetries for the $\pi^{+}n$ reaction. 
The photon beam asymmetries are sensitive to the existence 
of nucleon resonances. 
Basically, the asymmetries are +1 for the $\rho$-exchange and 
are -1 for the $\pi$-exchange in the $t$-channel~\cite{Guidal}. 
Large positive asymmetries measured by CEA, DESY, and SLAC suggest 
that $\rho$-exchange in the $t$-channel is the dominant reaction 
mechanism at $E_{\gamma}$ = 3.0 GeV~\cite{BarYam}, 
3.4 GeV~\cite{Burfeindt}, 
12 GeV~\cite{Schwitters}, 
and 16 GeV~\cite{Sherden,Quinn}. 
However, some theoretical models predict asymmetries with
large positive values in the case of 
$\pi$-exchange in the $t$-channel ~\cite{Guidal,Said}. 
The CLAS and GRAAL collaborations measured 
the asymmetries in a wide angular range without 
forward and backward $\pi^{+}$ angles for 
$E_{\gamma}$=1.102-1.862 GeV~\cite{Dugger2} and 
$E_{\gamma}$=0.8-1.5 GeV~\cite{Graal}, respectively. 
Asymmetry data in the LEPS kinematical region are missing 
in the world data set. 

%%%%%%% Regge model %%%%%%%%
Data at higher energies in the small $|t|$ region 
($|t|<$1 GeV$^{2}$/$c^{2}$) taken by SLAC 
were extensively studied by using
Regge models~\cite{Guidal,Sibirtsev,Yu}. 
The Regge models do not work correctly near the threshold region where 
the $s$-channel is dominant. 
It is questionable whether the Regge models work in the medium 
energy region. 
In the case of the $\gamma p$ $\rightarrow$ $K^{+}\Lambda$ reaction, 
Regge-Plus-Resonance model calculations successfully apply 
the Regge model at medium energies~\cite{Corthals,Ryckebusch}. 
The present LEPS data, which were taken in the small $|t|$ region 
and over a wide energy range, are suitable for 
studying the applicability of the Regge models. 
The Mandelstam variable $s$ dependence of the cross sections 
is expected to provide important information on 
the Regge trajectories exchanged in the 
$t$-channel, as demonstrated by Refs.~\cite{Pomeron,Boyarski,Chiu}. 
Testing the consistency between the results from 
the photon-beam asymmetries
and the cross sections helps us understand the $\pi^{+}$ 
photoproduction reaction. 

%%%%%%% End of introduction
Since the LEPS spectrometer~\cite{Nakano} was designed to efficiently detect 
a $\phi$ meson decaying to $K^{+}$ and $K^{-}$ in the forward angles, 
there were huge backgrounds of positrons and electrons. 
Using an Aerogel cherenkov counter was necessary to obtain 
clean $\phi$-meson production \cite{Mibe,Ishikawa,Chang1,Chang2,Chang3} and 
hyperon production 
\cite{Zegers,Sumihama,Kohri1,Niiyama,Hicks,Muramatsu2,Kohri2} 
data although high-momentum charged pion data were 
rejected by the online trigger. 
When the wavelength of the laser was changed from the UV to the deep-UV 
region, the photon beam intensity and trigger rate decreased. 
We took charged pion data for the first time in 2007. 
In this article, new LEPS data on differential cross sections and 
photon-beam asymmetries for the $\vec{\gamma} p$ $\rightarrow$ 
$\pi^{+}n$ reaction are presented.

\section{Experiment and data analysis}

The experiment was carried out by using the LEPS beam line \cite{Nakano} 
at the SPring-8 facility. 
The photon beam was produced by the laser backscattering technique 
using a deep-UV laser with a wavelength of 257 nm~\cite{Muramatsu}. 
The energy range of the tagged photon beam was from 1.5 to 2.96 GeV. 
The laser light was linearly polarized with a typical polarization 
degree of 98\%. 
The polarization of the tagged photon beams was 88\% 
at 2.96 GeV and was 28\% at 1.5 GeV. 
The photon beam was incident on a liquid hydrogen target (LH$_{2}$) with 
a length of 16 cm. 

Charged particles produced at the target were detected at 
forward angles using the LEPS spectrometer. 
Since the main purpose of the present experiment was to detect 
$K^{\ast 0}$ decaying to $K^{+}$ and $\pi^{-}$ 
with high momenta~\cite{Hwang,Hwang2}, 
the Aerogel cherenkov counter was not used. 
Electrons and positrons were effectively vetoed by installing a 
plastic scintillation counter at the downstream position of
the three drift chambers. 
The size of the scintillation counter was 40 mm in height, 
185 mm in width, and 20 mm in thickness.
The scintillation counter had a small hole 
20 mm in height and 50 mm in width 
that allowed the incident $\gamma$-ray beam to pass through. 
The details concerning the detector configuration and the 
quality of particle identification are given in 
Refs.~\cite{Nakano,Sumihama,Hwang2}. 

The $\pi^{+}$ meson events were identified from its measured mass 
within 3$\sigma$ where 
$\sigma$ is the momentum dependent mass resolution. 
The events of $\pi^{+}$-mesons generated in the LH$_{2}$ target were selected 
by the z-vertex distribution, 
and the contamination events from the start counter placed downstream from 
the target are 0.3\% at most. 

\begin{figure}[htb]
\includegraphics[height=10.3cm,width=8.5cm]{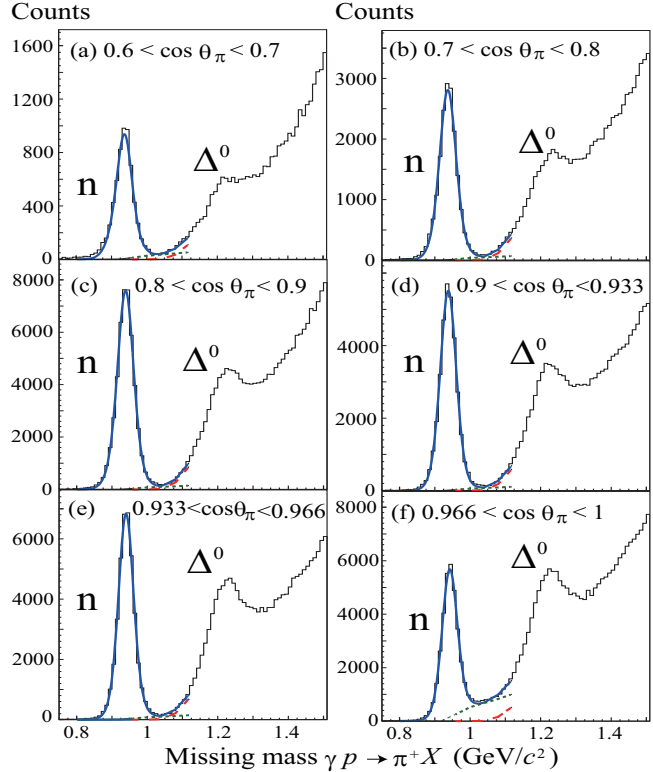}
\vspace*{-0.3cm}
\caption{\label{fig:miss} Missing mass spectra for the 
$\gamma p$ $\rightarrow$ $\pi^{+}X$ reaction for 
(a) 0.6$<$ $\cos\theta_{\pi}$ $<$0.7, 
(b) 0.7$<$ $\cos\theta_{\pi}$ $<$0.8, 
(c) 0.8$<$ $\cos\theta_{\pi}$ $<$0.9, 
(d) 0.9$<$ $\cos\theta_{\pi}$ $<$0.933, 
(e) 0.933$<$ $\cos\theta_{\pi}$ $<$0.966, and  
(f) 0.966$<$ $\cos\theta_{\pi}$ $<$1 with $E_{\gamma}$=1.5-2.95 GeV. 
The thick solid curves are the results of the fits, and the dotted and 
dashed curves are the contributions from the positron background and the 
$\pi\pi$ production events, respectively.}
\end{figure}

Figure \ref{fig:miss} shows the missing mass spectra for the 
$\gamma p$ $\rightarrow$ $\pi^{+}X$ reaction. 
Neutron peaks are clearly observed at 0.94 GeV/$c^{2}$ and 
bumps due to $\Delta^{0}$(1232) are also observed. 
The results for the $\pi^{+}\Delta^{0}$(1232) production are 
reported elsewhere~\cite{Kohri3,Kohri4}, although they are still preliminary.
The number of $\pi^{+}n$ events is about 171 k in total. 
The $\gamma p$ $\rightarrow$ $\pi^{+}n$ reaction events are selected 
by fitting the missing mass spectra with a Gaussian function for 
the neutron peak, a positron background curve, 
and a $\pi\pi$ production curve. 
The acceptance of the LEPS spectrometer for 
$\pi^{+}$-mesons is obtained by GEANT simulations. 

\newpage
\section{Results}

\subsection{Differential cross sections d$\sigma$/d$\cos\theta_{\pi}$}

\begin{figure}[h]
\resizebox{0.95\textwidth}{!}{\includegraphics{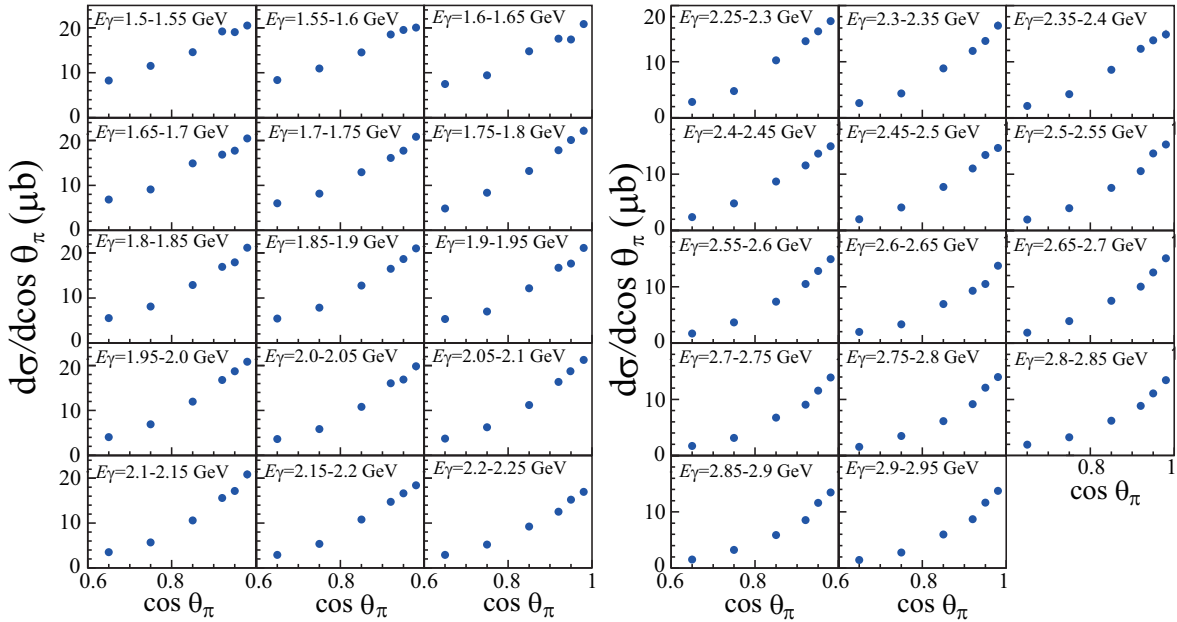}}
\caption{\label{fig:cross-ang} Differential cross sections 
d$\sigma$/d$\cos\theta$ 
for the $\gamma p$ $\rightarrow$ $\pi^{+}n$ reaction as a function 
of $\cos\theta_{\pi}$. }
\end{figure}

Figure \ref{fig:cross-ang} shows the differential cross sections 
for the $\gamma p$ $\rightarrow$ $\pi^{+}n$ reaction as a function 
of $\cos\theta_{\pi}$ in the center-of-mass frame. 
Systematic uncertainties of target thickness and photon flux 
are 1\% and 3\%, respectively. 
The cross sections increase rapidly as $\cos\theta_{\pi}$ approaches
1 in most of the energy regions.
The angular dependence is relatively small at around 
$E_{\gamma}$=1.5 GeV.
Forward peaking of the cross sections is observed, which 
suggests that there are significant $t$-channel contributions 
in the reaction mechanisms for this kinematical region. 
In the present work, we could not confirm the sharp rising of
the cross sections at very forward 
$\pi^{+}$ angles observed in the SLAC data~\cite{Boyarski}.

\begin{figure}[htb]
\includegraphics[height=9.5cm,width=8.5cm]{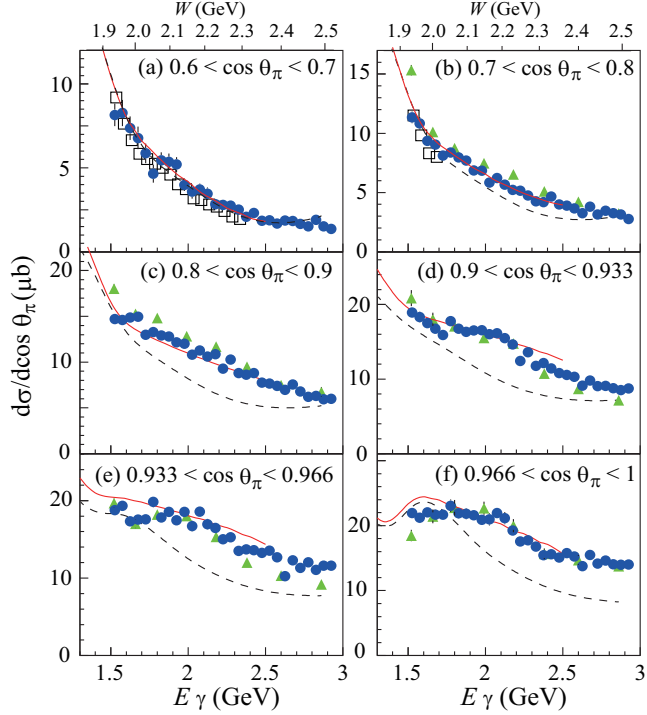}% 
\caption{\label{fig:cross-pipn} Differential cross sections for 
the $\gamma p$ $\rightarrow$ $\pi^{+}n$ reaction for 
(a) 0.6$<$ $\cos\theta_{\pi}$ $<$0.7, 
(b) 0.7$<$ $\cos\theta_{\pi}$ $<$0.8, 
(c) 0.8$<$ $\cos\theta_{\pi}$ $<$0.9, 
(d) 0.9$<$ $\cos\theta_{\pi}$ $<$0.933, 
(e) 0.933$<$ $\cos\theta_{\pi}$ $<$0.966, and 
(f) 0.966$<$ $\cos\theta_{\pi}$ $<$1 
with $E_{\gamma}$=1.5-2.95 GeV. 
The closed circles are the present LEPS data.
The open squares and the closed triangles are the CLAS~\cite{Dugger1} and
DESY data~\cite{Boschhorn1,Boschhorn2}, respectively. 
The solid curves are the results of the SAID analysis by the George
Washington University group~\cite{Said}. 
The dashed curves are the results of partial wave analysis by 
the Bonn-Gatchina group~\cite{BG}.}
\end{figure}

Differential cross sections as a function of $E_{\gamma}$ 
are shown in Fig.~\ref{fig:cross-pipn}. 
The cross sections monotonically decrease with increasing photon 
beam energy for 0.6$<$ $\cos\theta_{\pi}$ $<$0.9.
For 0.9$<$ $\cos\theta_{\pi}$ $<$1, the cross sections are almost constant
for $E_{\gamma}$=1.5-2.2 GeV($W$=1.9-2.2 GeV) and decrease above 
$E_{\gamma}$=2.2 GeV.
The constant cross sections are considered to be due to
a nucleon or $\Delta$ resonance at forward $\pi$ angles 
as reported by DESY~\cite{Boschhorn1,Boschhorn2}. 

The LEPS cross sections for the $\pi^{+}n$ reaction are in good
agreement with the CLAS~\cite{Dugger1} 
and DESY~\cite{Boschhorn1,Boschhorn2} cross sections. 
The SAID analysis~\cite{Said} reproduced the present data very well 
for $E_{\gamma}$ $<$2.5 GeV. 
The Bonn-Gatchina partial wave analysis calculations~\cite{BG} reproduce 
the present data well for 0.6$<$ $\cos\theta_{\pi}$ $<$0.8, 
but the calculations underestimate the data at small angles. 
The Bonn-Gatchina calculations were not fit to the DESY data, and 
the curves for $\cos\theta_{\pi}$ $>$ 0.7 are pure predictions. 

\subsection{Differential cross sections d$\sigma$/d$t$}

\begin{figure}[htb]
\includegraphics[height=11cm,width=6.5cm]{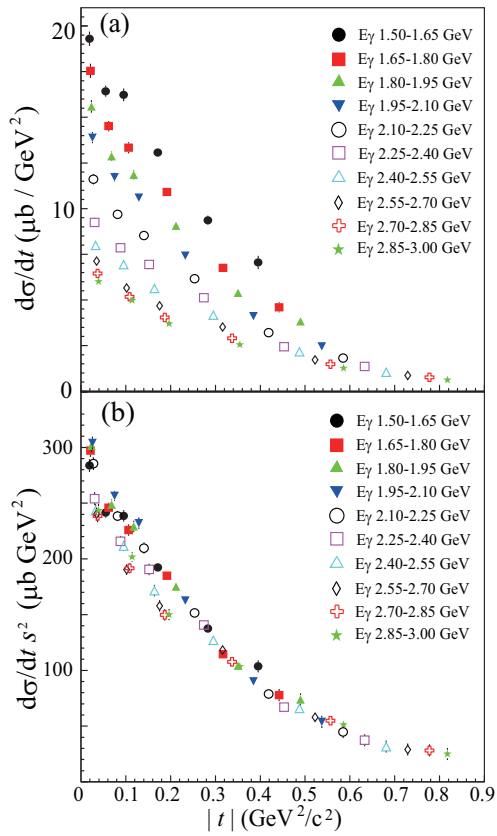}% 
\caption{\label{fig:scaling} (a) Differential cross sections d$\sigma$/d$t$ 
for the $\gamma p$ $\rightarrow$ $\pi^{+}n$ reaction as a function of 
$|t|$. 
(b) Differential cross sections scaled with $s^{2}$ as a function of 
$|t|$. }
\end{figure}

Figure \ref{fig:scaling}(a) shows differential cross sections 
d$\sigma$/d$t$ for the $\gamma p$ $\rightarrow$ $\pi^{+}n$ reaction 
as a function of $|t|$. 
With increasing photon energy, the cross sections decrease. 
Based on the Regge theory assuming a single trajectory, the $s$ 
dependence of the cross sections is written as 
\begin{equation}
\frac{d\sigma}{dt} = C(t)\Bigl(\frac{s}{s_{0}}\Bigr)^{2\alpha(t)-2}, 
\end{equation} 
where $C(t)$ and $\alpha(t)$ are functions of $t$ only, 
$s_{0}$ is a baryonic scale factor taken to be 1 GeV$^{2}$ and
$s$ is calculated as $s=M_{p}^{2}+2M_{p}E_{\gamma}$
with $M_{p}$ as the proton mass. 
The scaling of d$\sigma$/d$t$ with $s^{2}$ almost 
removes the energy dependence as shown in Fig.~\ref{fig:scaling}(b). 
This result suggests $\alpha(t)\approx$ 0.
A similar result was obtained by CLAS collaboration for
the $\gamma p$ $\rightarrow$ 
$K^{+}\Lambda$ reaction for 
$E_{\gamma}$=0.91-2.95 GeV~\cite{Bradford}. 

\begin{figure}[htb]
\includegraphics[height=12cm,width=6.5cm]{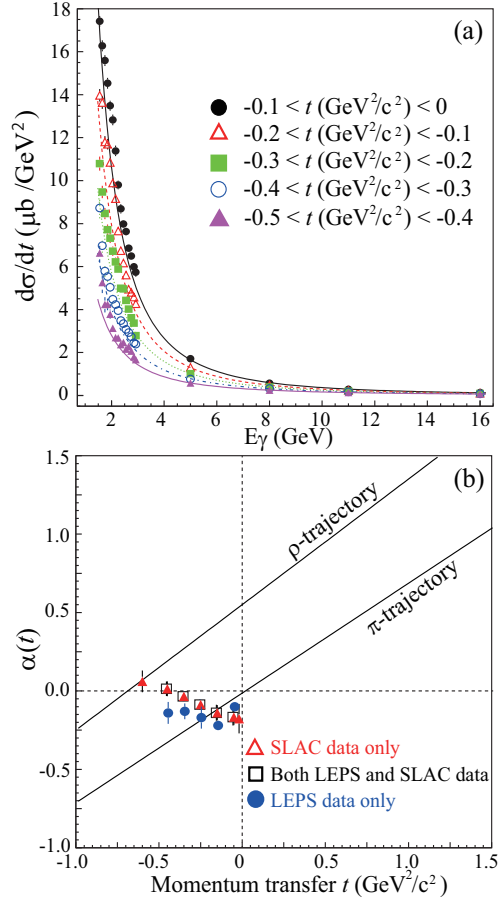}% 
\caption{\label{fig:alpha} (a) Differential cross sections d$\sigma$/d$t$ 
for the $\gamma p$ $\rightarrow$ $\pi^{+}n$ reaction as a function of 
$E_{\gamma}$. 
The data $E_{\gamma}<$3.0 GeV were measured by LEPS and the data 
$E_{\gamma}\geq$5 GeV were measured by SLAC. 
The curves are the results of exclusive fits to the SLAC data. 
(b) The $\alpha$(t) values for the $\gamma p$ $\rightarrow$ $\pi^{+}n$ reaction 
are the results from the SLAC(triangle), the LEPS and SLAC(squares), 
and the LEPS(circles). 
The $\pi$ and $\rho$ trajectories are represented using the functions of
$\alpha_{\pi}(t)$=0.7($t-m^{2}_{\pi}$) 
and $\alpha_{\rho}(t)$=0.55+0.8$t$, respectively~\cite{Guidal}. }
\end{figure}

A small energy dependence still remains in the small $|t|$ region 
for $E_{\gamma}>$2.4 GeV in Fig.~\ref{fig:scaling}(b).
The assumption of $\alpha(t)\approx$ 0 does not work well. 
Further studies are necessary to obtain effective $\alpha(t)$ values
which give information on 
what trajectory is effective in the $\gamma p$ $\rightarrow$
$\pi^{+}n$ reaction. 
Figure \ref{fig:alpha}(a) shows the differential cross sections 
d$\sigma$/d$t$ for the $\gamma p$ $\rightarrow$ $\pi^{+}n$ reaction 
measured by LEPS and SLAC. 
The cross sections were fit with the function 
$C(t)s^{2\alpha(t)-2}$, 
where $C(t)$ and $\alpha(t)$ are free parameters for 
each $t$. 
Each curve is a result of fitting exclusively to the SLAC data, which 
were measured at high energies and are considered to be dominated 
by $t$-channel contributions. 
The curves slightly underestimate the LEPS data. 

The effective $\alpha(t)$ values are shown in Fig.~\ref{fig:alpha}(b). 
The $\alpha(t)$ values obtained from the SLAC data, the LEPS and SLAC 
data, and the LEPS data are close to each other. 
The present cross sections measured for $E_{\gamma}$=1.5-2.95 GeV 
are found to have almost the same $s$ dependence as the SLAC data. 
The $\alpha(t)$ values obtained from the LEPS data are slightly 
smaller than those from the SLAC data 
for $t<$-0.1 GeV$^{2}$/c$^{2}$. 
The differences of the $\alpha(t)$ values are considered to come from 
the differences of reaction mechanisms between the LEPS data and
the SLAC data. 
Differences between the LEPS data and the curves in 
Fig.~\ref{fig:alpha}(a) are about 10-20\% on average and
estimated to be due to resonance contributions in the $s$-channel. 
The resonance contributions are small and the $t$-channel 
contributions are dominant in the LEPS kinematical region. 
The application of the Regge theory to the LEPS kinematical region 
seems to be acceptable. 
The $\alpha(t)$ values range between -0.22 and 0.06. 
The $s$ dependence of the cross sections at $t$ close to 
0 GeV$^{2}$/c$^{2}$ favors the single $\pi$ trajectory, while the 
dependence at $t$ close to -0.5 GeV$^{2}$/c$^{2}$ cannot be 
simply explained by the single $\pi$-trajectory. 

\subsection{Photon-beam asymmetry}

\begin{figure}[htb]
\includegraphics[height=8.8cm,width=7.5cm]{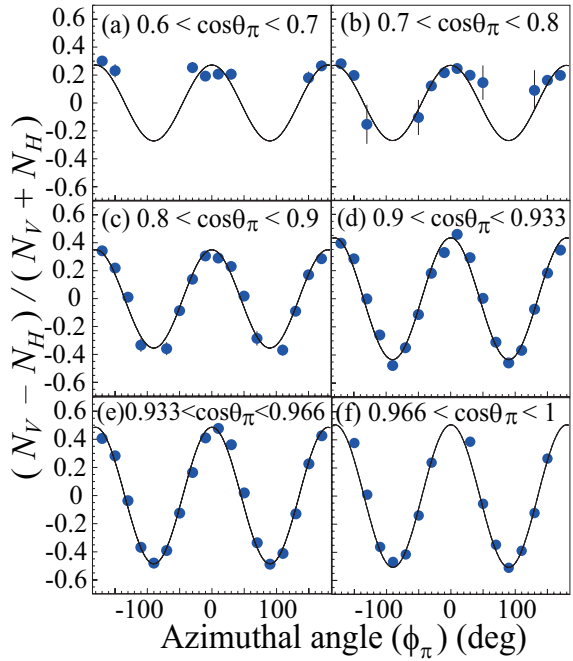}%
\caption{\label{fig:azimuthalangle} The ratio 
($N_{V}-N_{H}$)/($N_{V}$+$N_{H}$) as a function 
of $\pi^{+}$ azimuthal 
angle ($\phi_{\pi}$) for the $\vec{\gamma} p$ $\rightarrow$ $\pi^{+}n$
reaction for 
(a) 0.6$<$ $\cos\theta_{\pi}$ $<$0.7, 
(b) 0.7$<$ $\cos\theta_{\pi}$ $<$0.8, 
(c) 0.8$<$ $\cos\theta_{\pi}$ $<$0.9, 
(d) 0.9$<$ $\cos\theta_{\pi}$ $<$0.933, 
(e) 0.933$<$ $\cos\theta_{\pi}$ $<$0.966, and 
(f) 0.966$<$ $\cos\theta_{\pi}$ $<$1 with $E_{\gamma}$=1.5-2.9 GeV.
The solid curves are the result of the fits with
$P_{\gamma}\Sigma \cos2\phi_{\pi}$.}
\end{figure}

We have measured the $\vec{\gamma}p$ $\rightarrow$ $\pi^{+}n$ data using
vertically and horizontally polarized photon beams. 
The photon-beam asymmetry $\Sigma$ is given as 
\begin{equation}
P_{\gamma}\Sigma \cos2\phi_{\pi} = \frac{N_{V} - N_{H}}{N_{V} + N_{H}}, 
\end{equation}
where $N_{V}$ and $N_{H}$ are the $\pi^{+}$ yields with 
vertically and horizontally polarized photon beams, respectively, 
after correcting the difference of photon counts in both polarizations. 
$P_{\gamma}$ is the polarization of the photon beams and 
$\phi_{\pi}$ is the $\pi^{+}$ azimuthal angle. 
Figure \ref{fig:azimuthalangle} shows the ratio 
($N_{V}-N_{H}$)/($N_{V}$+$N_{H}$) for the $\vec{\gamma} p$ $\rightarrow$
$\pi^{+}n$ reaction events for $E_{\gamma}$=1.5-2.9 GeV. 

Since the LEPS spectrometer has a wide acceptance for the horizontal 
direction and a narrow acceptance for the vertical direction, 
the number of events is small at $\phi_{\pi}$=$\pm$90$^{\circ}$ for 
0.6$<$ $\cos\theta_{\pi}$ $<$0.9. 
On the other hand, the number of events is small 
at $\phi_{\pi}$=$\pm$0$^{\circ}$ 
and $\pm$180$^{\circ}$ for 0.966$<$ $\cos\theta_{\pi}$ $<$1 because 
the veto counter for removing $e^{+}e^{-}$ was installed. 
The ratio ($N_{V}-N_{H}$)/($N_{V}$+$N_{H}$) is 
large at 0$^{\circ}$ and $\pm$180$^{\circ}$ and small 
at $\pm$90$^{\circ}$, so 
$\pi^{+}$-mesons prefer to scatter at $\phi_{\pi}$ angles perpendicular 
to the polarization plane. 
The photon-beam asymmetries are found to be positive. 
The amplitude of the ratio increases as the polar angle ($\theta_{\pi}$) 
of the $\pi^{+}$-mesons becomes smaller. 

\begin{figure}[htb]
\includegraphics[height=11cm,width=8.5cm]{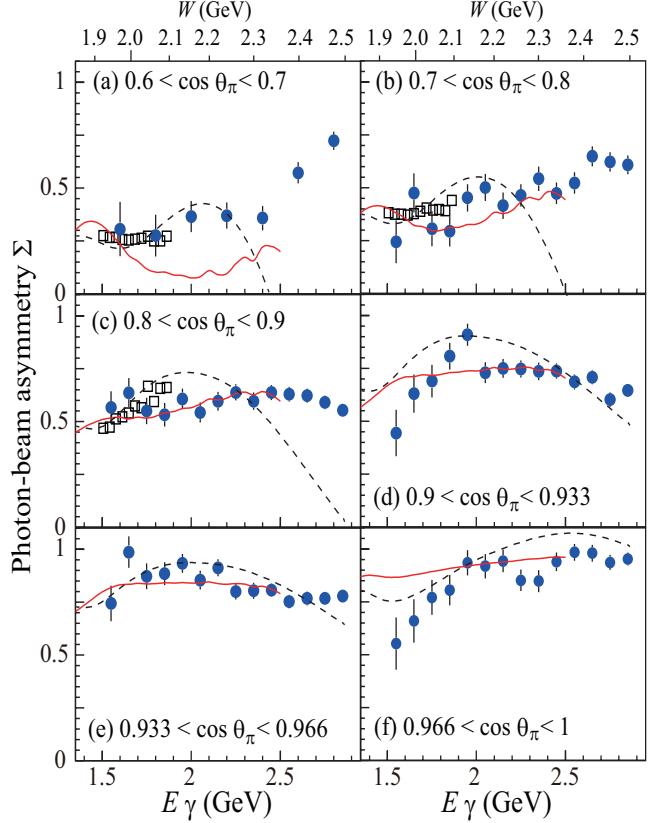}% 
\caption{\label{fig:asymmetry} Photon-beam asymmetries for 
the $\vec{\gamma} p$ $\rightarrow$ $\pi^{+}n$ reaction 
for $E_{\gamma}$=1.5-2.95 GeV. 
The closed circles are the present LEPS data and
the open squares are the CLAS data~\cite{Dugger2}. 
The solid curves are the results of the SAID analysis by
the George Washington University 
group~\cite{Said}.
The dashed curves are the results of partial wave analysis performed
by the Bonn-Gatchina group~\cite{BG}.}
\end{figure}

Figure \ref{fig:asymmetry} shows the photon-beam asymmetries for the 
$\vec{\gamma} p$ $\rightarrow$ $\pi^{+}n$ reaction. 
The systematic uncertainty of the measurement of the laser 
polarization is $\delta\Sigma$=0.02. 
The effects of the positron contamination in the $\pi^{+}$ sample 
and the start counter contamination in the LH$_{2}$ target are 
removed. 
The asymmetries are positive in all the LEPS kinematical region, 
which can be explained by $\rho$-meson exchange in the $t$-channel. 

The photon-beam asymmetries are small at large $\pi^{+}$ angles, while 
the asymmetries become large and approach unity 
at small $\pi^{+}$ angles. 
It is interesting that this angular dependence is different 
from the asymmetries obtained for 
the $\vec{\gamma} p$ $\rightarrow$ $K^{+}\Lambda$ and
$K^{+}\Sigma^{0}$ reactions. 
The asymmetries for those two reactions become small 
at small $K^{+}$ angles~\cite{Zegers}. 
The asymmetries for 0.9$<$ $\cos\theta_{\pi}$ $<$1 and $E_{\gamma}$=1.5-2 GeV 
are slightly smaller than those at higher energies. 
The differential cross sections also have different energy dependence 
in this kinematical region as shown in Fig.~\ref{fig:cross-pipn}. 
These results might suggest the existence of a nucleon or $\Delta$ 
resonance although the final conclusion should wait until a 
partial wave analyses is done over a wide kinematical region. 

The agreement between the LEPS data and the CLAS data is good although 
the overlap of the photon energy region is limited.
The SAID analysis by the George Washington University group well 
reproduces the present data for 0.7$<\cos\theta_{\pi}<$0.966 
and $E_{\gamma}<$2.5 GeV. 
The SAID analysis underestimates the present data for 
0.6$<\cos\theta_{\pi}<$0.7. 
Calculations by the Bonn-Gatchina partial wave analysis almost 
reproduce the present data for $E_{\gamma}<$2.4 GeV.
The Bonn-Gatchina calculations underestimate the present data for 
$\cos\theta_{\pi}<$0.9 and $E_{\gamma}>$2.4 GeV.
The calculations are pure predictions for $E_{\gamma}>$2.4 GeV. 

The result of the $\rho$-exchange for the positive asymmetries
seems to be in contradiction to the result obtained 
from the Regge model studies shown in Fig.~\ref{fig:alpha} where 
the $\pi$-trajectory almost explains the $s$ dependence 
of the cross sections d$\sigma$/d$t$ in the small $\pi$ 
angles($t$ close to 0 GeV$^{2}$/c$^{2}$). 
The theoretical calculations given in Ref.~\cite{Said,Guidal}
predict positive photon-beam asymmetries in the case of the $\pi$-exchange. 
The positive asymmetries are obtained by an interference 
between the $\pi$-exchange in the $t$-channel 
and the $s$-channel resonances. 

\begin{figure}[htb]
\includegraphics[height=13cm,width=8.5cm]{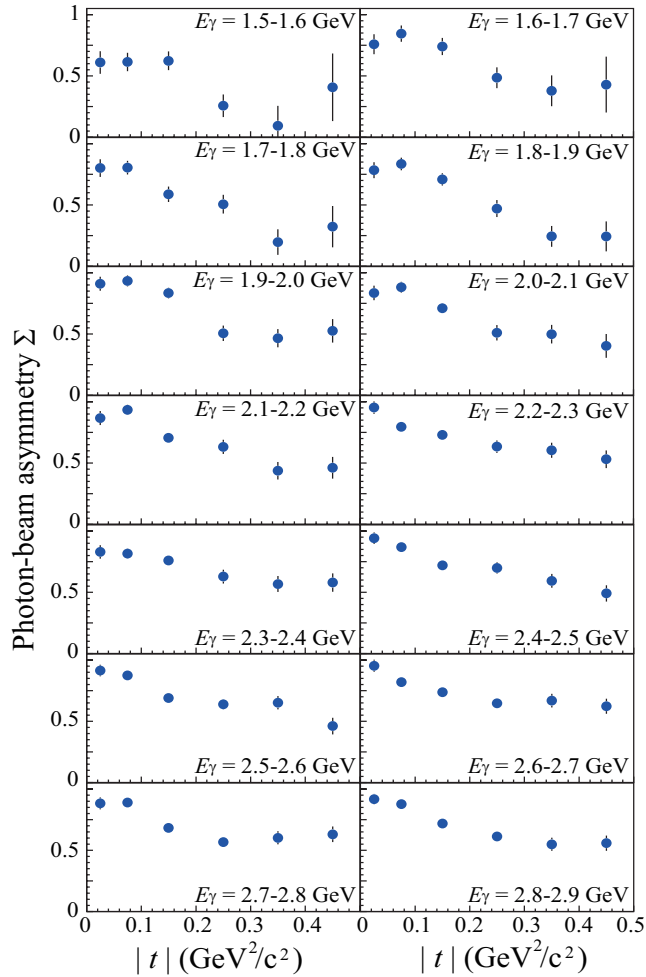}% 
\caption{\label{fig:asym-tdep} Photon-beam asymmetries for 
the $\vec{\gamma} p$ $\rightarrow$ $\pi^{+}n$ reaction 
as a function of $|t|$. }
\end{figure}

Figure \ref{fig:asym-tdep} shows photon-beam asymmetries
for the $\pi^{+}n$ reaction as a function of $|t|$. 
The asymmetries become large as $|t|$ becomes smaller. 
A similar $|t|$ dependence is observed at 16 GeV~\cite{Sherden}. 
The $\rho$-exchange contribution becomes small 
in the small $|t|$ region~\cite{Guidal}. 
The forward peaking asymmetry observed in Fig.~\ref{fig:asym-tdep} 
cannot be explained by a $\rho$-exchange contribution. 
Large positive asymmetries in the small $|t|$ region 
could be due to $\pi$-exchange interference 
with the $s$-channel~\cite{Guidal}. 
A final conclusion needs further advancements in theory 
or new data observables which can distinguish 
between the two contributions. 

\section{Summary}

We have carried out a photoproduction experiment observing 
the $\vec{\gamma} p$ $\rightarrow$ $\pi^{+}n$ reaction 
by using linearly 
polarized tagged photon beams with energies from 1.5 to 2.95 GeV. 
Differential cross sections and photon-beam asymmetries 
have been measured for 0.6$<$ $\cos\theta_{\pi}$ $<$1. 
The differential cross sections monotonically decrease as the 
photon beam energy increases for 0.6$<$ $\cos\theta_{\pi}$ $<$0.9, while 
the cross sections are close to constant values 
for $E_{\gamma}$=1.5-2.2 GeV ($W$=1.9-2.2 GeV) and 
decrease above $E_{\gamma}$=2.2 GeV for 0.9$<$ $\cos\theta_{\pi}$ $<$1. 
This energy dependence for $E_{\gamma}$=1.5-2.2 GeV is inferred
to be due to a nucleon or $\Delta$ resonance although 
the final conclusion should wait for a partial wave analyses over 
a wider kinematical region. 

Regge model studies on the $s$ dependence of d$\sigma$/d$t$ give 
$\alpha(t)$ values close to the $\pi$-trajectory at $t$ close to 
0 GeV$^{2}$/c$^{2}$. 
Positive asymmetries found for 
the $\vec{\gamma} p$ $\rightarrow$ $\pi^{+}n$ reaction 
can be explained by $\rho$-exchange in the $t$-channel. 
%Consistent pictures are not obtained to understand the measured results 
%of the cross sections and asymmetries. 
Large positive asymmetries in the small $|t|$ region could be 
explained by the $\pi$-exchange interference with the $s$-channel 
as suggested by some theoretical calculations~\cite{Guidal,Said}. 
Experimentally, we are developing a polarized HD target~\cite{Kohri5} 
for LEPS experiments, and CLAS has already taken data with 
polarized butanol~\cite{Ritchie,Akbar} and HD targets~\cite{Ho}. 
Rich physics observables measured by using polarized targets and polarized 
photon beams are expected to appear soon. 
Theoretically, partial wave analyses using these physics 
observables are available. 
The photon-beam asymmetry is a strong constraint to 
theoretical models. 
Our data will provide an important contribution to
advanced theoretical studies that we hope will clarify
the hadron photoproduction dynamics in the near future. 

\begin{acknowledgments}
The authors gratefully acknowledge the staff of 
the SPring-8 facility for the supports with excellent 
experimental conditions. 
The experiments were performed at the BL33LEP of SPring-8 
with the approval of the Japan Synchrotron Radiation Research 
Institute (JASRI) as a contract beamline (Proposal No. BL33LEP/6001). 
H.K. thanks Prof. H. Kamano and Prof. T. Sato 
for fruitful discussions and Prof. R. Workman 
for providing results of the SAID analysis. 
H.K. also thanks Prof. Ulrike Thoma, Prof. Eberhard Klempt,
and Prof. Andrey Sarantsev for providing results
and useful information on the Bonn-Gatchina calculations. 
This research was supported in part by 
the Ministry of Education, Science, Sports and Culture of Japan, 
the National Science Council of the Republic of China, 
the National Research Foundation of Korea, and 
the U.S. National Science Foundation. 
\end{acknowledgments}

\nocite{*}

\bibliography{apssamp}% Produces the bibliography via BibTeX.

\end{document}